\documentclass{iopart}
\usepackage{natbib}

\usepackage[pdftex]{graphicx}
\usepackage{moreverb}
\usepackage[utf8]{inputenc}

\begin{document}

\title{A Light-Weight Communication Library for Distributed Computing}

\author{Derek Groen$^{1,2}$, Steven Rieder$^{1,2}$, Paola Grosso$^2$, Cees de Laat$^2$, Simon Portegies Zwart$^1$}

\address{$^1$ Leiden Observatory, Leiden University, P.O. Box 9513, 2300 RA Leiden, The Netherlands}
\address{$^2$ University of Amsterdam, Amsterdam, the Netherlands}
\ead{djgroen@strw.leidenuniv.nl}

\maketitle
\begin{abstract}
We present MPWide, a platform independent communication library for
performing message passing between computers.  Our library
allows coupling of several local MPI applications through a long distance network
and is specifically optimized for such communications. The implementation is
deliberately kept light-weight, platform independent and the library can be
installed and used without administrative privileges.  The only
requirements are a C++ compiler and at least one open port to a wide
area network on each site. In this paper
we present the library, describe the user interface, present
performance tests and apply MPWide in a large scale
cosmological N-body simulation on a network of two computers, one in
Amsterdam and the other in Tokyo.
\end{abstract}

\section{Introduction} 
A parallel application can run concurrently on multiple supercomputers
provided one is able to coordinate the tasks between them and limit
the performance overhead of the wide area communications. The
advantage of using a distributed infrastructure lies in the enormous
amounts of storage, RAM and computing performance it makes available.
Distributed computing therefore allows us to solve large scale
scientific problems~\cite{hoekstra2008}. Starting from the coupling of
Intel Paragons over an ATM network~\cite{paragon} in the early 1990s,
distributed parallel applications have become very popular.

An efficient method to program a parallel application is the Message Passing
Interface (MPI \cite{mpi}), a language-independent communication protocol that
coordinates the computing tasks in parallel programs. MPI is often used for
intra-site parallelization, but it can also be used for message passing in a
distributed infrastructure. In this case, processes exchange data with their
local peers, as well as processes at other sites. Prior efforts in the use of MPI on distributed
infrastructures are abundant~\cite{Hockney,Manos,1467242} and several implementations have
emerged which support execution across sites~\cite{openmpi,mpich-g2}. With 
respect to $N$-body simulations Gualandris et. al.~\cite{Gualandris2007} have 
demonstrated that it is possible to use grid-enabled clusters of PCs connected via regular
internet, grid middleware and MPICH-G2~\cite{mpich-g2}. However, the vast majority of
MPI implementations require all participating nodes to have public IP
addresses, which is generally undesirable for supercomputer environments
for security reasons. Furthermore these implementations do not have a built-in
optimization to fully exploit dedicated network circuits, a central
component in multi-supercomputer infrastructures.



The lack of flexibility in deployment and link-specific optimizations of
grid-oriented MPI implementations in distributed supercomputer environments led
us to develop MPWide, a light-weight socket library specially aimed for
wide-area message passing between supercomputers. In this paper we present
several performance results and apply MPWide to parallelize a large-scale
cosmological $N$-body simulation across two supercomputers.

\section{Related work} 

Several grid message-passing libraries and frameworks have been
developed with the intent to make distributed computing possible between
sites that have restrictive firewall policies.
PACX-MPI \cite{pacx-mpi} is specifically geared for parallelization
across sites and does not require compute nodes to have a public IP address.
Instead, it forwards inter-site communications through two forwarding demon processes on
each site. Such a setup works reasonably well for applications that
have been parallelized over multiple supercomputers using regular
internet~\cite{Stewart2003}, but the two communication process restriction is less
optimal when using multiple sites in a dedicated network environment.
The Interoperable MPI (IMPI) \cite{IMPI} standard has also been designed to
specifically facilitate execution across sites, but at the time
of writing very few of the vendor-tuned implementations on supercomputers
support IMPI. Also, IMPI requires the installation of a centralized and globally
accessible server and does not support path-specific optimizations.

NetIbis \cite{NetIbis} and PadicoTM \cite{PadicoTM} are two communication
frameworks which are able to establish connections using bootstrap links, thus
not requiring public IP addresses. However, PadicoTM also requires the use of a
centralized rendez-vous node for bootstrapping, and thereby some means of centralized
connectivity. Both Ibis \cite{Ibis} and NetIbis are sufficiently flexible to use in a restricted
supercomputer environment, but introduce a communication overhead compared to regular
socket communications. These libraries are therefore less suitable for high-performance
message passing over dedicated inter-supercomputer networks.


\section{Architecture of MPWide} \label {sec:arch}

\subsection{Design} 

MPWide is a light-weight communication library which connects multiple
applications on different supercomputers, each of them running with the
locally recommended MPI implementation. It can be installed by a local user
without administrative privileges, has a very limited set of software
requirements, and the application programmer interface is similar to that of
MPI. The applications are deployed separately for each supercomputer, and 
use MPWide to connect with each other upon startup. We are considering to 
add support for automated deployment, but to accomplish this we require a 
method to initiate applications on remote sites. The development of such a 
mechanism is not straightforward, because the access and security policies 
tend to be different for each supercomputer.

MPWide has been designed to facilitate message passing between supercomputers 
and construct/modify custom communication topologies. The MPWide library is 
linked to the application at compile time and requires only the 
presence of UNIX sockets and a C++ compiler. MPWide provides an abstraction
layer on top of regular sockets with methods to construct a 
communication topology, to adjust the parameters of individual communication 
paths and to perform message passing and forwarding across the topology.
MPWide does not link against local MPI implementations, but can be used
to combine multiple programs parallelized with MPI. Maintaining separate
implementations for intra- and inter-site message passing makes it easier
to specifically optimize and debug long-distance communication paths while
relying on well-tested and vendor-tuned software for optimal intra-site 
communication performance.

\subsubsection{Data transport in the wide area network}
Dedicated network circuits are excellently suited for facilitating
data transport between supercomputers. The highly deterministic bandwidths of
optical circuits (or \textit{lightpaths}) reduce the communication
time while properly tuned transport layer protocols increase the
application throughput in the absence of competing traffic.

During the development of MPWide, we have examined the communication
performance of several protocols by transferring data between two nodes in the
Netherlands using a 10 Gbps optical network that was looped via Chicago, USA.
We ran tests using both the TCP and the UDP network protocols. Plain UDP does
not ensure the integrity of data packets however, and is therefore
unsuitable for message passing. As an alternative, we instead tested the performance 
of two modified UDP implementations which feature mechanisms to ensure data integrity. These are
Reliable Blast UDP (RBUDP \cite{rbudp}, which is part of the Quanta toolkit
\cite{He2003919}) and the UDP-based Data Transfer protocol (UDT \cite{udt}).
The tests using TCP were run with both a single communication stream and with
multiple streams in parallel. 

We achieved a network throughput of less than 1 Gbps using RBUDP or UDT, and a
throughput of up to 6 Gbps using parallel TCP streams. A full technical report
on these preliminary performance tests can be found in \cite{CGtests}. Based on
these results we decided to rely on multiple streams with a TCP-based protocol.
This is a well-known and proven techniques to improve network performance in
the WAN \cite{TCP-perf-Qiu}.

\subsubsection{Functionality and programming interface}
In MPWide, the communication takes place through \emph{channels}. Each
channel makes use of a single socket and provides a bidirectional 
connection between two ports on two hosts. On network paths where
the use of parallel TCP streams provides a performance benefit, it is
possible to use multiple channels concurrently on the same path. The message
passing and forwarding functions in MPWide are designed to 
operate concurrently on multiple channels when needed.

Channels are locally defined at initialization and may be closed, 
modified and reopened at any time during execution. This allows us 
to alter the communication topology at run-time, for example to restart 
or migrate part of the MPWide-enabled application. 

Once one or more communication channels have been established, the
user can transfer data using the communication calls in the
MPWide API. Table \ref{tab:MPW-calls} provides an overview of the
functionality provided by MPWide.

\begin{table}[!th]
    \centering
    \begin{tabular}{|l|l|l|}

\hline

command name  & functionality \\

\hline

{\tt MPW\_Barrier()}   & Synchronize between two ends of the network.             \\

{\tt MPW\_Cycle()}     & Send buffer over one set of channels, receive from other.\\  

{\tt MPW\_DSendRecv()} & Send/receive buffers of unknown size using caching.      \\  

{\tt MPW\_Init()}      & Set up channels and initialize MPWide.                   \\

{\tt MPW\_Finalize()}  & Close channels and delete MPWide buffers.                \\

{\tt MPW\_Recv()}      & Receive a single buffer (merging the incoming data).     \\

{\tt MPW\_Relay()}     & Forward all traffic between two channels.                \\  

{\tt MPW\_Send()}      & Send a single buffer (splitted evenly over the channels).\\

{\tt MPW\_SendRecv()}  & Send/receive a single buffer.                            \\

\hline

      \end{tabular}
  \caption{List of MPWide function calls. In addition to this list, each function 
    has a variant call with a prefix 'P' which operates on one send and/or recv 
    buffer per channel.}
  \label{tab:MPW-calls}
\end{table}

Since message passing can be performed over multiple channels in parallel, it
is possible to communicate with multiple hosts simultaneously. For example, 
the user can scatter data across multiple processes with a single 
{\tt MPW\_Send()} call or gather data from multiple hosts with a single 
{\tt MPW\_Recv()}. Each function has a variant call with a prefix 'P' 
(e.g., {\tt MPW\_PSend()})
which takes an array of buffer pointers instead of one buffer pointer. These 
functions use one pointer for each channel, and the size of each separate buffer 
can be explicitly specified. Consequently, {\tt
  MPW\_PSend()} or {\tt MPW\_PRecv()} functions can be used to
respectively scatter and gather data which is not equally distributed
across the hosts.

Both {\tt MPW\_Cycle()} and {\tt MPW\_DSendRecv()} also support the receiving of data
buffers which are of unknown size (but not larger than a given size
limit provided by the user). This feature may provide some performance
improvement in long distance environments at the expense of possible
excessive memory consumption, as separate calls to exchange data size
information are no longer required

An MPWide code example is shown in Fig.\ref{fig:code_snippet}. There
we initialize MPWide with one single-stream path to a local network
address, the {\tt LANChannels}, as well as a double stream path to a
different site, the {\tt WANChannels} (lines 1-12). The program then
initializes two buffers (line 14-15); it reads 100 bytes of data from
the local connection with {\tt MPW\_Recv()} (line 18); it exchanges
this data with the remote WAN communication node using {\tt
  MPW\_SendRecv()} (line 20). At this point, the program has received
the data from the remote WAN node, and forwards this data to the local
connection (line 22).

\begin{figure}

\begin{listing}{1}
int NumChannels = 3; // Total number of channels.
int NumLAN      = 1; // Number of LAN channels.
int NumWAN      = 2; // Number of WAN channels.
int MsgSize     = 100;

int Hosts = {"10.0.0.100","123.45.67.89", "123.45.67.89"};
int Ports = {6000, 6001, 6002};

MPW_Init(Hosts, Ports, NumChannels);

int LANChannels[NumLAN] = {1};
int WANChannels[NumWAN] = {2,3};

char* SendBuf = new char[MsgSize];
char* RecvBuf = new char[MsgSize];

// Recv from LAN.
MPW_Recv    (SendBuf, MsgSize, LANChannels, 1);                   
// WAN exchange.
MPW_SendRecv(SendBuf, MsgSize, RecvBuf, MsgSize, WANChannels, 2); 
// Send to LAN.
MPW_Send    (RecvBuf, MsgSize, LANChannels, 1);                   

/* ( ... Process data and delete SendBuf and RecvBuf. ... ) */

MPW_Finalize();
\end{listing}
\caption{Example code of the MPWide functionality}
\label{fig:code_snippet}
\end{figure}

\subsection{Forwarder}
When running an application across multiple sites, the processes on
one site are not always directly able to communicate with the other
site. In many cases this problem can be resolved by forwarding the
messages to intra-cluster communication nodes, which do have access to
all other sites through the wide area (dedicated) network. However,
when the application uses a topology containing multiple dedicated
networks, it will be necessary to forward messages from one network to
another. The {\tt MPW\_Relay()} function provides such message forwarding
for MPWide channels, and has been incorporated into the MPWide
Forwarder. The Forwarder provides message forwarding for MPWide in
user space, connecting an MPWide channel from one network to that of
another. It can therefore serve as relay process between nodes that
are otherwise unable to contact each other or be put on intermediate
nodes on very long network lines to mitigate the performance impact
of packet loss. This latter method has been implemented and applied 
previously in the Phoebus project \cite{Phoebus}.

\subsection{Implementation} 

We implemented MPWide using C++ in combination with GNU C sockets and POSIX
threads \cite{pthreads}. MPWide creates and destroys threads on the fly
whenever a communication call is made. With modern kernels, the overhead
of creating and destroying threads is very small, and using MPWide we 
were able to reach nearly 10 Gigabit per second (Gbps) with message passing tests over local 
networks. For longer network paths, the high latency results in 
an even smaller relative overhead for thread creation/destruction. 
We have considered creating threads only at startup and managing them at runtime, 
but these modifications would increase the complexity of the code and only 
offer a limited performance benefit, as threading overhead plays a marginal 
role in wide area communication performance.

Aside from the ability to hardwire each communication, the library also
supports a number of customizable parameters:

\begin{itemize}
\item Number of concurrent streams for each communication call.
\item Data feeding pace of sending and receiving.
\item TCP window size for each individual socket.
\end{itemize}
The maximum number of streams and the TCP window size may be
restricted by local system policies. However, we were able to 
use up to 128 streams on most systems without requiring 
administrative rights. The code has been packaged and is 
publicly available at \\
{\tt http://castle.strw.leidenuniv.nl/software/mpwide.html}.


\section{Benchmarking MPWide} \label{sec:setup}

We have run a series of tests to measure the performance of
MPWide between two local supercomputer nodes, as well as two nodes connected by a 10 Gbps international network connection. 

For the local tests, we use two nodes of the Huygens supercomputer in
Amsterdam, the Netherlands \cite{Huygens}, where the nodes are connected by 8
parallel Infiniband links, each of which supports a maximum bandwidth of 20
Gbps. Our local tests use one out of these 8 Infiniband links. Each run consists of 100
two-way message exchanges, where we record the average throughput and the standard
error. First we performed 8 different tests using messages of 8 MB and
respectively 1, 2, 4, 8, 16, 32, 64 and 124 TCP streams in parallel. Due to
system limitations, we were unable to perform tests using more than 124
streams on this particular site. We then repeated the same series of runs with
message sizes of 64 and 512 MB.

The national tests were carried out between two sites of the Distributed ASCI Supercomputer 3 (DAS-3 \footnote{DAS-3: http://www.cs.vu.nl/das3/}),
one at the University of Amsterdam and one at the Delft University of 
Technology. Both sites are connected to regular internet with a 10 Gbps 
interface from the head node, and with a 1 Gbps interface from each compute
node. A detailed specification can be found in columns 4 and 5 of 
Table \ref{tab:sc-specs}. We performed the tests using the system default TCP
window sizes (16 kB send and 85 kB recv).

For the international tests, we performed the same series of message exchanges,
but now using one Huygens node and one node of the Louhi supercomputer in
Helsinki, Finland \cite{Louhi}, which are both connected to the DEISA shared network
with a 10 Gbps interface. The round trip time of this network between
Huygens and Louhi is 37.6 ms and we applied a TCP window size of 16 MB. 
The specifications of both supercomputers can be found in columns 2 and 3 
of Table \ref{tab:sc-specs}. 

\begin{table}[!h]
    \centering
    \begin{tabular}{|l|l|l|l|l|}

\hline

 & Huygens
 & Louhi & DAS-3 Ams & DAS-3 Delft\\
\hline

CPU vendor           & IBM        & Cray    & AMD         & AMD\\

Architecture         & Power6     & XT4     & Opteron     & Opteron\\

Number of nodes      & 104        & 1012    & 41          & 68\\

Cores per node       & 32         & 4       & 4           & 2\\

CPU frequency [GHz]  & 4.7        & 2.3     & 2.2         & 2.4\\

Memory per core [GB] & 4/8        & 1/2     & 1           & 2\\

\hline

      \end{tabular}
  \caption{Specifications of the Huygens and Louhi supercomputers, as well as the two sites of the DAS-3 Dutch Grid.}
  \label{tab:sc-specs}
\end{table}


\subsection{Results}\label{sec:results}
\subsubsection{Local tests} 

\begin{figure}[h!]

  \centering
  \includegraphics[scale=0.3]{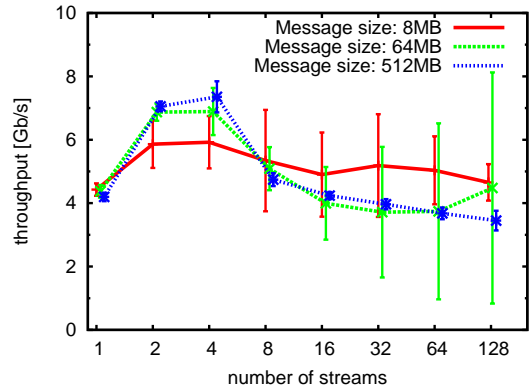}
  \caption{Measured throughput in Gbps as a function of the number of communications streams used between two nodes on Huygens. The throughput is given for runs with 1 to 124 threads and message sizes of respectively 8, 64 and 512 MB.}

\label{Fig:LocalThreads}
\end{figure}

The results of the local performance tests, performed in March 2009, are found
in Fig.~\ref{Fig:LocalThreads}. The local network line has a very low latency
($< 0.1$ ms) and is therefore quickly saturated when using multiple streams. In
our results we found an increase in throughput when using 2 or 4 streams, but
using more concurrent streams results in a performance decrease. When
increasing the number of streams, the overhead caused by creating and
destroying threads also increases, and may have contributed to this performance
loss. However, if this were the case, we would observe a much steeper decline
in performance for 8 MB messages than for 512 MB messages, as these
communications take less time overall, and are thus more easily dominated
by threading overhead. We therefore conclude that this overhead is caused by
saturation of the local network line. The maximum throughput achieved in these
tests is close to the theoretical maximum bandwidth of 10 Gbps. This proves
that the MPWide library can efficiently utilize the available bandwidth, if
optimal settings are used.

\subsubsection{National tests}

\begin{figure}[h!]

  \centering
    \includegraphics[scale=0.3]{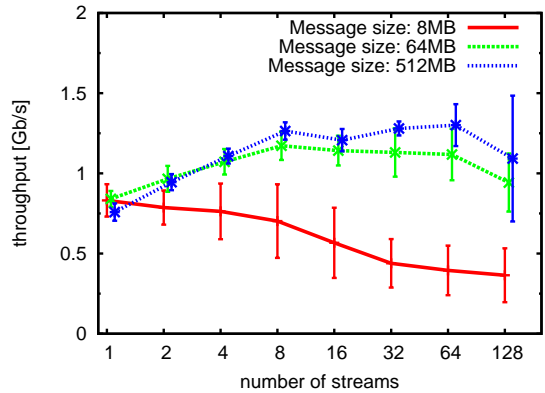}
      \caption{Measured throughput in Gbps as a function of the number of
      communications streams used between the DAS-3 site in Amsterdam and the
      DAS-3 site in Delft. The throughput is given for runs with 1 to 128
      threads and message sizes of respectively 8, 64 and 512 MB.}

\label{Fig:DAS3Threads}
\end{figure}

We carried out the national tests over two sites of the DAS-3 Dutch Grid. One
site resides at the University of Amsterdam and the other site is located at
the Delft University of Technology. The round-trip time of the path between 
Amsterdam and Delft is 2.1 ms. The results of these tests are found in
Fig.~\ref{Fig:DAS3Threads}. Although the tests used the regular
internet backbone the fluctuations in our measurements are limited. When
exchanging messages of 8 MB size, we obtain the best performance using a single
stream, as the use of additional streams results in a lower and more variable 
performance. This is caused by the fact
that message passing performance over multiple streams is limited by the slowest
streams. For larger message sizes, however, using a single stream does not
result in an optimal performance. Instead, we find that the best results are
obtained using 8 streams (for 64 MB) to 32 streams (for 512 MB). Although a
high peak performance is obtained when using 64 or more streams, the
sustained performance is lower because the excess streams can cause network
congestion. The round-trip time of 2.1 ms did not significantly reduce the 
achieved throughput in our tests.

\subsubsection{International tests}

\begin{figure}[h!]

  \centering
  \includegraphics[scale=0.3]{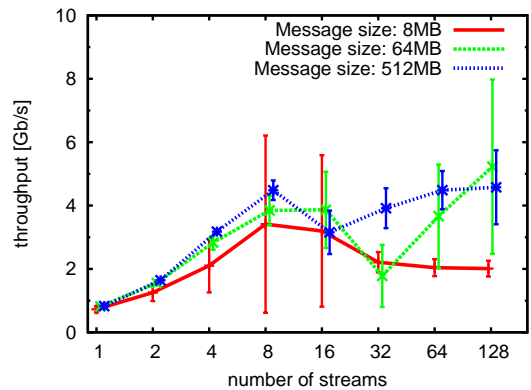}    
  \caption{Measured throughput in Gbps as a function of the number of communication streams used between Huygens and Louhi. The throughput is given for runs using 1 to 124 streams and message sizes of respectively 8, 64 and 512 MB.}

\label{Fig:HuygensLouhiThreads}
\end{figure}

We show the results of the international tests between Louhi and Huygens,
performed in March 2009, in Fig.~\ref{Fig:HuygensLouhiThreads}. The tests were
performed over a shared 10 Gbps network with frequent background network
traffic. To minimize the impact of this background traffic, we performed our
tests during a quiet period of the day. However, a few of our tests had
background interference, causing fluctuations in the measured throughput. When
exchanging 8 MB messages, the throughput rate no longer increases once we scale
beyond 8 parallel streams. Here, the throughput rate is limited to about 3.5
Gbps due to the high network latency and the small message size. For message
sizes of 64 MB and especially 512 MB, the network latency no longer constrains
the achieved throughput rate. As a result, we achieved a higher throughput when
using more streams. Similar to the national tests, we notice larger
fluctuations in performance for larger message sizes. The highest average
throughput we achieved was about 4.64 Gbps, which we achieved using 64 streams
and a message size of 512 MB.

\section{Testing performance in a production environment}\label{sec:cg-desc}


We originally developed MPWide to manage the long-distance message
passing in the CosmoGrid project \cite{CosmoGrid}. CosmoGrid is a
large-scale cosmological project which aims to perform a
dark matter simulation of a cube with sides of 30 Mpc using
supercomputers on two continents.  In this simulation, we use the
cosmological $\Lambda$ Cold Dark Matter model
\cite{1981PhRvD..23..347G} which defines a constant fraction of the
overall energy density for dark energy to model the accelerating
expansion of the universe. We apply this model to simulate the dark
matter particles with a parallel tree/particle-mesh $N$-body
integrator, GreeM \cite{Ishiyama09}. This integrator can be run either 
as a single MPI application, or as multiple MPI applications on different
supercomputers. In the latter case, the wide area communications are 
performed using MPWide. We use GreeM to calculate
the dynamical evolution of $2048^3$ ($\sim 8.590$ billion) particles
over a period of time from redshift $z=65.35$ to $z=0$. More
information about the parameters used and the scientific rationale can
be found in \cite{CosmoGrid}. 

\begin{figure}[t!]

  \centering
  \includegraphics[scale=0.3]{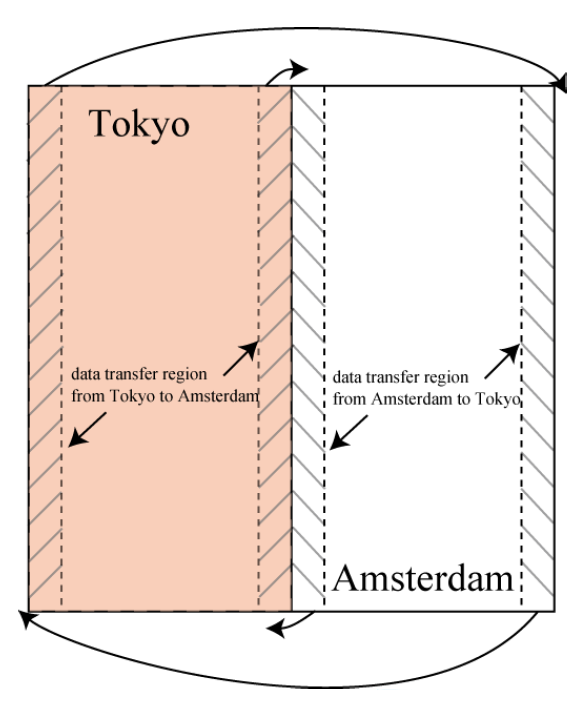}    
  \caption{Data decomposition overview of the CosmoGrid simulation
    when run on two supercomputers \cite{CosmoGrid}.}

\label{Fig:DataTopology}
\end{figure}

Before the simulation is launched, the initial condition is decomposed
in slices for each site, and in blocks within that slice for each
process. Each block contains an equal number of particles but may
vary in volume. A simulation process loads one block during startup,
and calculates tree and particle mesh force interactions at every
step. These force calculations require the exchange of particles
with neighboring processes (and sites, see
Fig.\ref{Fig:DataTopology}). These force calculations require the exchange of particles
with neighboring processes (and sites, see Fig.\ref{Fig:DataTopology})
as well as the exchange of mesh cells. In addition, a number of smaller 
communications are performed to balance the load across all processes.

We have used GreeM together with MPWide in a set of test runs,
which consist of full-lengths simulations of a limited scale ($256^3$ 
particles). Also, we have performed a run across two supercomputers 
which consists of a limited part of the production 
simulation described earlier.

\subsection{Test experiments}

We have run three test simulations, of which each one uses a different 
infrastructure. All runs were carried out over two sites, with 30 calculation
processes and one communication process per site. We performed one run  
on the DAS-3 testbed and one run across the Huygens and Louhi supercomputers. 
Both infrastructures are described in section ~\ref{sec:setup}, and for both 
infrastructures we carried out simulations with communication over 1 TCP stream, 
as the average data volume is only a few MB per communication.

For the third run we have used the Huygens supercomputer 
in combination with a Cray XT-4 supercomputer
located at the Center for Computational Astrophysics in Tokyo,
Japan. The Cray XT-4 consists of 740 nodes which run on a quad-core
2.2GHz AMD Opteron and have 8GB RAM each. To exchange data between the
sites we reserved and used a 10 Gbps dedicated light path in the GLIF
network\cite{GLIF}, which has a round trip time of 273 milliseconds.
This run was performed prior to the other two runs, using an older
version of the code and the library. Unfortunately we were unable to
reserve the lightpath for a new test run using our improved setup.
For this test we used 64 concurrent TCP streams.

\begin{figure}[t!]

  \centering
  \includegraphics[scale=0.5]{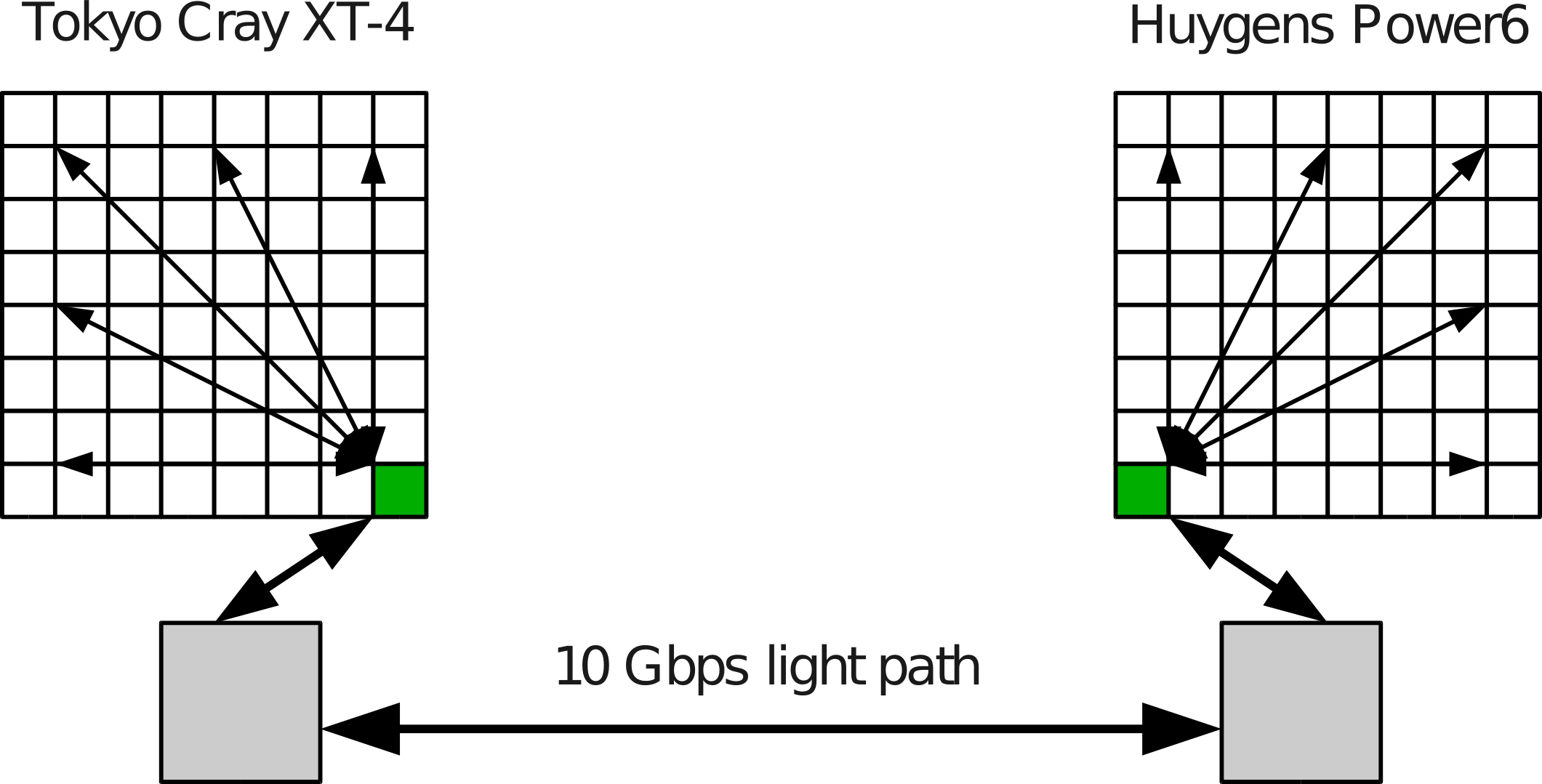}    
  \caption{Network topology example of the CosmoGrid simulation when
    run on two supercomputers, one in Amsterdam, the Netherlands and
    one in Tokyo, Japan. Data transfers within the local supercomputer
    are performed using MPI (thin arrows), whereas other
    communications are performed using MPWide (thick arrows). The
    communication nodes (indicated by the gray boxes) reside outside
    of the MPI domains, and therefore use MPWide for all
    communications. Before the data is transferred to the communication 
    node, it is gathered on a central process on the local supercomputer 
    (indicated by the green boxes).}

\label{Fig:NetworkTopology}
\end{figure}

A detailed overview of the communication topology during the
simulation can be found in Fig.\ref{Fig:NetworkTopology}. Each of the
supercomputers has been equipped with one specialized communication
node. These nodes are each connected to the high-speed local
supercomputer network and are linked together by the 10 Gbps light
path. MPWide is used to transfer the locally gathered data to the
communication node, forward it to the other site using the light path,
and finally to deliver the data to the remote MPI simulation.

\subsubsection{Results on DAS-3 Dutch grid}

\begin{figure}[t!]

  \centering
  \includegraphics[scale=0.3]{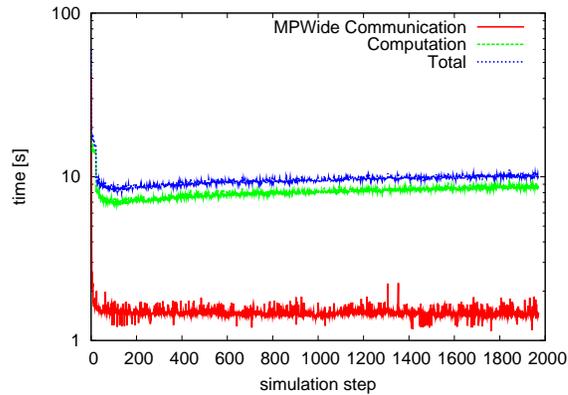}
  \caption{Measured wall-clock time spent (in log-scale) on each
    simulation step for a $256^3$ particle test run on the DAS-3
    between Amsterdam and Delft. The full-length run was
    performed using 62 cores, with 30 cores residing on each
    supercomputer and 2 cores used for communication only. The top dotted line indicates total time spent, the
    dashed line indicates time spent on calculation and the bottom solid
    line represents time spent on communication with MPWide.}
\label{Fig:256Das3}
\end{figure}

The performance results of our test simulation on the DAS-3 can be found
in Fig.~\ref{Fig:256Das3}. Here we find that the simulation performance is 
dominated by calculation, with a communication overhead less than
20 percent of the overall wallclock time throughout the run. As
we used regular internet for the wide area communication, our simulation
performance is subject to the influence of background network traffic.
The two performance dips which can be found around step 1300 and 1350 are 
most likely caused by incidental increases in background traffic.

\subsubsection{Results on Amsterdam and Helsinki supercomputers}

\begin{figure}[t!]

  \centering
  \includegraphics[scale=0.3]{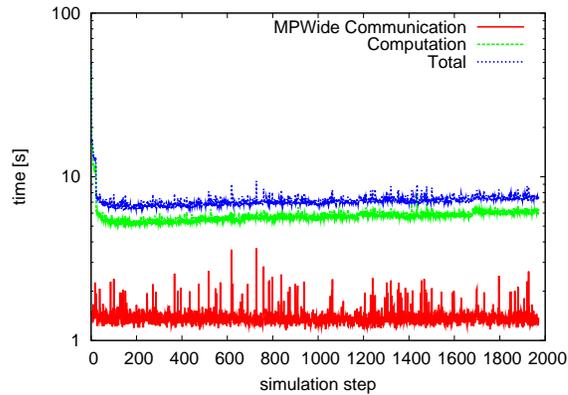}
  \caption{Measured wall-clock time spent (in log-scale) on each
    simulation step for a $256^3$ particle test run on the DEISA network
    between Amsterdam and Helsinki. See Fig.~\ref{Fig:256Das3} for an explanation
    of the lines.}
  \label{Fig:256LH}
\end{figure}

The performance results of our test simulation between Amsterdam and Helsinki 
are shown in Fig.~\ref{Fig:256LH}. The obtained performance is similar to that
on the DAS-3, although two differences can be noted. First, the calculation time
is $\sim 25$ percent lower due to the superior performance of the supercomputer
nodes. Second, although the average communication performance is similar to that
observed on the DAS-3, we observe more variability in the communication performance.
We are at this point uncertain about the exact nature of this variability. The 
DEISA network is shared with other institutions, so the presence of background
traffic may have decreased our communication performance.

\subsubsection{Results on Amsterdam and Tokyo supercomputers}\label{sec:oct2008test}

\begin{figure}[t!]

  \centering
  \includegraphics[scale=0.3]{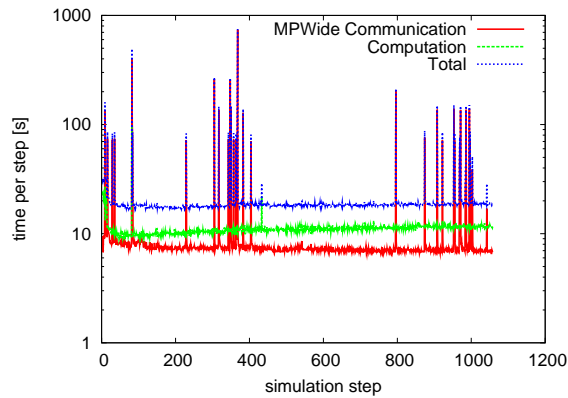}    
  \caption{Measured wall-clock time spent (in log-scale) on each
    simulation step for the $256^3$ particle test run. See Fig.~\ref{Fig:256Das3} for an explanation
          of the lines.}
\label{Fig:256}
\end{figure}

The run between Huygens and the Tokyo Cray-XT4 was carried out in October 2008,
before any of the other experiments in this paper, and served as a dress
rehearsal for both the Tree-PM simulation code and MPWide. The simulated
problem consisted was of equal size of the previous simulations and uses
the same number of processes. However, we performed the run using an older
version of the code and different initial condition files.  The
performance results of this run can be found in Fig.~\ref{Fig:256}. During
this test run, the time spent on calculation is roughly constant throughout
the run, with a peak occurring during startup and a few points where
snapshots are written. The time spent in communication is generally lower
than the calculation time, taking about 7 to 10 seconds per step. However,
we also observe a number of communication performance drops. These
temporary decreases in performance were almost exclusively caused by single 
communications stalling for an extended period, which in turn were caused 
by periods of packet loss.

\subsection{Production}
We have executed a production-sized simulation between Amsterdam and
Tokyo to measuring the performance of the code when it is used for 
production. Based on the results described in Sec.~\ref{sec:oct2008test}, we 
made a few changes to the network settings before performing the second run. 
We disabled the TCP memory suppression mode, increased the TCP window 
sizes, and increased the {\tt sysctl} queue limit for upper-layer processing.
Due to the limited length of our network reservation, we were not able
to test the effects of modifying each of these settings in detail.

\begin{figure}[t!]

  \centering
  \includegraphics[scale=0.3]{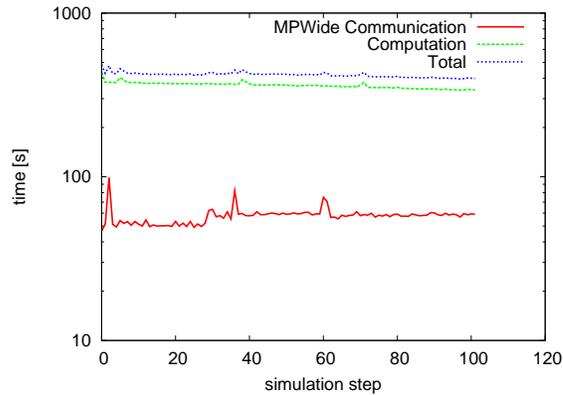}    
  \caption{Measured wall-clock time (in log-scale) for each simulation
    step for a partial $2048^3$ particle run. The run, which uses some
    adjusted TCP settings, was performed using 750 cores, with 500
    cores used on Huygens and 250 cores used on the Tokyo Cray. An
    explanation of the lines can be found in the caption of
    Fig.~\ref{Fig:256Das3}.}

\label{Fig:2048}
\end{figure}

The production-sized run was performed on 752 cores in total. 
The topology of this run was asymmetric,
using 500 cores on Huygens and 250 cores on the Cray for
calculation. The full run lasted just under 12 hours, during which we performed 102
simulation steps. The performance results of this run can be found in
Fig.~\ref{Fig:2048}. In this full-scale run, the calculation time
dominated the overall performance, and was slightly higher at
startup and during steps where snapshots were written. The
communication performance is generally more constant than in the small-scale run between
Amsterdam and Tokyo , with
fewer and less severe performance drops and a slight increase in time 
after step 30 in the simulation. This increase may have been caused by the TCP
buffering sizes, which the local system may change during
run-time. Overall, the total communication time per step was between
50 and 60 seconds for most of the simulation, and constituted about
one eighth of the total execution time.

\section{Conclusions and future work} \label{sec:future-concl}

We present MPWide, a communication library to perform message passing between
supercomputers. MPWide provides message passing that is intrinsically
parallelized, and can be used for high-performance computing across
multiple supercomputers. The library allows for customization of individual
connections and has a light-weight design, which makes it well-suited for 
connecting different supercomputer platforms. We have shown results
from local and wide area performance tests, and applied MPWide to combine
two MPI applications into a very large parallel simulation across several
wide area compute infrastructures. During our tests, we reached a sustained
throughput of up to 4.64 Gbps over a long-distance 10 Gbps network. In addition,
we were able to run an $N$-body simulation across two continents with $2048^3$
particles. During this simulation, about one eighth of the execution time was
spent on communications.

Given that the parallel application is sufficiently scalable (which is the case
for the $N$-body integrator used in this work), MPWide can be used to efficiently 
parallelize production applications across multiple supercomputers. Future efforts
to improve the usability of MPWide may include the integration with debugging tools and
visualization toolkits, the introduction of group communicators and collective 
operations (similar to {\tt MPI\_COMM\_WORLD} in MPI implementations), and the addition of
an automatic deployment mechanism.

\section*{Acknowledgements} \label{sec:ack}
We would like to thank Jun Makino for his valuable help, input and
 hospitality during the development of this library. Also we are
grateful to Tomoaki Ishiyama for his work on interfacing and running
the TreePM code with MPWide and his valuable feedback during
development. We also would like to thank Hans Blom for highly useful
discussions and performing preliminary tests.
Performing the intercontinental simulations would not have been
possible without the help of Keigo Nitadori, Steve McMillan, Kei
Hiraki, Stefan Harfst, Walter Lioen, Petri Nikunen, Ronald van der
Pol, Mark van der Sanden, Peter Tavenier, Huub Stoffers, Alan Verlo,
Joni Virtanen and Seiichi Yamamoto.
This research is supported by the Netherlands organization for
Scientific research (NWO) grant \#639.073.803, \#643.200.503 and \#643.000.803, the
European Commission grant for the QosCosGrid project (grant number:
FP6-2005-IST-5 033883), SURFNet with the GigaPort project, NAOJ, the International Information Science
Foundation (IISF), the Netherlands Advanced School for Astronomy
(NOVA), the Leids Kerkhoven-Bosscha fonds (LKBF) and the Stichting
Nationale Computerfaciliteiten (NCF). We thank the DEISA Consortium 
(www.deisa.eu), co-funded through the EU FP6 project RI-031513 and 
the FP7 project RI-222919, for support within the DEISA Extreme 
Computing Initiative (GBBP project). 

\bibliographystyle{unsrt}
\bibliography{citations}

\end{document}